# Investigating Coordination of Hospital Departments in Delivering Healthcare for Acute Coronary Syndrome Patients using Data-Driven Network Analysis

Tesfamariam M Abuhay[1,3*], Yemisrach G Nigatie[3], Oleg G Metsker[1],

Aleksey N Yakovlev[1,2], Sergey V Kovalchuk[1]

[1] ITMO University, Saint Petersburg, Russia
[2] Almazov National Medical Research Centre, Saint Petersburg, Russia
[3] University of Gondar, Gondar, Ethiopia
`tesfamariam.m.abuhay@gmail.com, yemisrach.getinet@uog.edu.et,`
`olegmetsker@gmail.com, yakovlev_an@almazovcentre.ru,`
`sergey.v.kovalchuk@gmail.com`

**Abstract.** Healthcare systems are challenged to deliver high-quality and efficient care. Studying patient flow in a hospital is particularly fundamental as it demonstrates effectiveness and efficiency of a hospital. Since hospital is a collection of physically nearby services under one administration, its performance and outcome are shaped by the interaction of its discrete components. Coordination of processes at different levels of organizational structure of a hospital can be studied using network analysis. Hence, *this article presents a data-driven static and temporal network of departments*. Both networks are directed and weighted and constructed using seven years' (2010-2016) empirical data of 24902 Acute Coronary Syndrome (ACS) patients. The ties reflect an episode-based transfer of ACS patients from department to department in a hospital. The weight represents the number of patients transferred among departments. As a result, the underlying structure of network of departments that deliver healthcare for ACS patients is described, the main departments and their role in the diagnosis and treatment process of ACS patients are identified, the role of departments over seven years is analyzed and communities of departments are discovered. The results of this study may help hospital administration to effectively organize and manage the coordination of departments based on their significance, strategic positioning and role in the diagnosis and treatment process which, in-turn, nurtures value-based and precision healthcare.

---

[*] Corresponding author

# 1 Introduction

Provision of healthcare is one of the fundamental expenditures and political agendas of every government. Norway, Switzerland and the United States are the world's three biggest healthcare spenders – paying per person $9,715 (9.6% of GDP), $9,276 (11.5% of GDP), and $9,146 (17.1% of GDP), respectively [1]. However, healthcare systems are challenged to deliver high-quality and efficient care because of aging population, epidemic and/pandemic (e.g., COVID-19), scarcity of resources and poor planning, organization and management of healthcare processes [2][3][4].

As a well-coordinated and collaborative care improves patient outcomes and decreases medical costs [5], there is a need for effective organization of operational processes in a hospital. Besides to this, recent healthcare delivery system reforms such as value-based healthcare, accountable care and patient-centered medical homes require a fundamental change in providers' relationship to improve care coordination [6].

Acute Coronary Syndrome (ACS) is an umbrella term for an emergency situations where the blood supplied to the heart muscle is suddenly blocked [7][8]. According to World Health Organization (WHO) [9], an estimated 17.9 million people died from Cardiovascular Diseases (CVDs) in 2016, representing 31% of all global deaths. Of these deaths, an estimated 7.4 million were due to ACS.

Whenever people feel symptoms of ACS, they visit a hospital as emergency or planned patient. During their stay in a hospital, ACS patients may move from one department to another department to get medical treatment or patients' laboratory samples and/or medical equipment may move from department to department.

Since hospital is a system that combines inter-connected and physically nearby services [10], its behavior and outcome are shaped by the interactions of its discrete components [11][12]. In other words, departments in a hospital deliver different but interdependent services. An output of one department's operation can be an input and/or precondition for one or many departments in the diagnosis and treatment processes. This makes the underlying processes of a hospital highly dynamic, interconnected, complex, ad hoc and multi-disciplinary [5][13][14].

Coordination of operational processes at different level of organizational structure of a hospital can be conceptualized, studied and quantified using graph/network analysis or Social Network Analysis (SNA) [6][15].

In organizational behavior studies, SNA can be defined as a set of social entities, such as people, groups, and organizations, with some relationships or interactions between them [16][17]. SNA allows to model, map, characterize and quantify topological properties of a network, discover patterns of relations and identify the roles of nodes and sub-groups within a network [16][18].

A systematic reviews [6][15][19][20][21][22] mentioned that SNA can be applied in healthcare setting to study interaction of healthcare professionals such as physician-nurse interactions and physician-physicians communication; diffusion of innovations, including adoption of medical technology, prescribing practices, and evidence-based medicine; and professional ties among providers from different organizations, settings, or health professions.

Chambers et al [19] mentioned that 50 of 52 studies used survey or observation to collect data for constructing and studying complex networks in healthcare setting. Nowadays, healthcare administrative data have been widely used to demonstrate SNA [5]. According to [5][6] and [12], the standard practice to construct network of healthcare providers (healthcare professionals) is based on "patient-sharing" concept meaning; two providers are considered to be connected to one another if they both deliver care to the same patient. To do so, first, bipartite network of patient-physician should be created. Next, this bipartite network can be projected to unipartite network of physicians only, where the ties reflect patient-sharing between physicians. For instance, Soulakis et al [5] made an attempt to visualize and describe collaborative electronic health record (EHR) usage for hospitalized patients with heart failure by creating 2 types of networks: the first is a directed bipartite network which represents interactions between providers and patient records and the second network is undirected and depicts shared patient record access between providers. In 2018, Onnela et al [12] have compared standard methods for constructing physician networks from patient-physician encounter data with a new method based on clinical episodes of care using data on 100% of traditional Medicare beneficiaries from 51 regions for the years 2005–2010.

However, hospital is a combination of departments and each department consists both human and material resources including medical equipment. As a result, effectiveness and efficiency of a hospital depends on organization, collaboration and availability of both human and material resources. Due to this, network of human resources (healthcare professionals) only may not reflect the real or complete picture of the underlying structure of the diagnosis and treatment process in a hospital. To the best of our knowledge, no one has studied collaboration of hospital departments using network analysis.

This article, therefore, investigates data-driven network of departments to answer the following research questions: *what is the underlying structure of network of departments that deliver healthcare for ACS patients? what are the main departments and their role in the diagnosis and treatment process of ACS patients? does the role of departments change over time? can we detect communities of departments which are highly interconnected?*

Answering these questions may give insight about the underlying organizational structure of departments, the role, strategic positioning and influence of departments in the diagnosis and treatment process and the interaction among departments and sub-groups of departments (communities). This would help to effectively structure and manage the collaboration among departments and sub-group of departments by maintaining the functionality, capacity and geographical proximity of departments according to their role and significance in the diagnosis and treatment process of ACS patients.

As fraction of seconds matter a lot in diagnosing and treating ACS patients, the results of this study may also help to optimize the operational processes and minimize time, cost and effort of patients and health professionals.

The rest of the paper is organized as follows: Section 2 outlines methods including data collection and preprocessing; Section 3 discusses results obtained and Section 4 presents conclusion and future works.

## 2    Methods

This study was conducted in collaboration with the Almazov National Medical Research Centre[Φ], a major scientific contributor and healthcare provider that delivers high-tech medical care. The study was reviewed and approved by the Institute Review Board (IRB) of the National Center of Cognitive Research (NCCR) at ITMO University.

Seven years', from 2010 to 2016, empirical data of 24902 ACS patients was collected from this hospital. The patient identifiers were excluded from the dataset to protect the privacy of patients. The event log data describes movement of patients from department to department in the center and the corresponding timestamps. Those departments visited by ACS patients from 2010 to 2016 are included in this study.

Each patient's event log data was sorted based on an event date. In the dataset, there are two IDs, patient-ID and episode-ID, which uniquely identify a patient and clinical episode of a patient, respectively. The patient-ID is constant over the life time of a patient, whereas the episode-ID changes as the clinical episode of a patient changes. Patient-ID has been commonly used to construct network of providers [6]. However, one patient may have many episodes in different time period and departments providing care to a patient in the context of one clinical episode may not be directly connected to another clinical episode [12]. In this case, the standard approach based on patient-ID would produce a network that does not correspond to the real connections. In this study, therefore, episode-ID was used and network of departments was constructed based on the chronological transfer or flow of ACS patients. The proposed approach does not require creating bipartite network of patient-provider and changing (projecting) the resulting network to unipartite network of providers only.

Two types of unipartite network was constructed, the first one is the static network and the second one is temporal network with one-year time window from 2010 to 2016. In order to reduce potential noises, the departments with less than 10 interaction (the number of patients transferred between departments) in seven years and less than 5 interaction in one year were excluded from the static and the temporal networks, respectively.

The networks were constructed as directed graph, where nodes represent departments and ties reflect the flow or transfer of patients from one department to another department. For instance, if a patient had a surgery and transferred to Intensive Care Unit (ICU), it forms a directed graph from surgery (source) to ICU (target). The number of patients transferred between departments was employed as a weight. i.e., the proposed network is both directed as well as weighted graph. Finally, Gephi 0.9.2 [23] was employed to visualize the structure of the network and generate both network and node level statistics.

---

## 3 Results

### 3.1 Static Network of Departments

The static network of departments contains 227 nodes that represent departments and 4305 edges that represent flow or transfer of patients. Both degree and weighted degree distributions (see Fig. 1) are positively skewed with a large majority of departments having a low degree and a small number of departments having a high degree. This shows that network of departments are scale free [16] meaning; there are few departments that are highly connected to other departments in the diagnosis and treatment process of ACS patients.

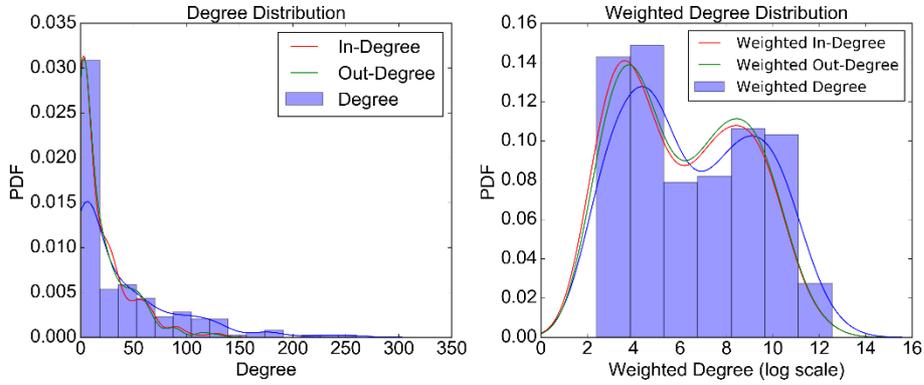

**Fig. 1.** Degree and weighted degree distribution of static network of departments.

The average degree and weighted degree account for 19 and 5800, in that order. This means that one department has an interaction with 19 departments on average by transferring or receiving 5800 patients on average over seven years. Even though the network of departments is sparse with density equals to 0.1, the average path length is short which accounts for 2.3. i.e., a given department may reach other departments in the network with 2.3 hops on average. Out of 227 departments in the static network, 27 departments are strongly connected meaning; 27 departments are connected to each other by at least one path and have no connections with the rest of the network.

**Table 1.** Top five departments with high support and influence.

| Departments | Degree | In-Degree | Out-Degree | Weighted-Degree | Weighted-In-Degree | Weighted-Out-Degree |
|---|---|---|---|---|---|---|
| Regular Laboratory | 260 | 131 | 129 | 282260 | 141055 | 141205 |
| Emergency Laboratory1 | 240 | 124 | 116 | 277469 | 138259 | 139210 |
| Functional Diagnostic | 219 | 105 | 114 | 171371 | 85964 | 85407 |
| Cardiology One | 182 | 93 | 89 | 101915 | 52381 | 49534 |
| Cardiology Two | 182 | 90 | 92 | 96148 | 49525 | 46623 |

When we see the strategic positioning of departments in the static network, Regular Laboratory department, Emergency Laboratory1 department and Functional Diagnostic department are receiving more requests as well as sending more results to other departments (see Table 1).

In addition to this, cardiology departments, ICU departments and heart surgery departments are also vital in the diagnosis and treatment processes of ACS patients. This shows that ACS requires intensive diagnosis and treatment procedures and care. These departments are significant in terms of giving support to and influencing activities of other departments. Therefore, maintaining the functionality, capacity and geographical proximity of these departments is vital so as to deliver effective and efficient healthcare for ACS patients.

These departments are also fundamental in connecting communities of departments as they have high betweeness and closeness centrality in the network (see Table 2). In addition to bridging regions of the network, they may also facilitate the information flow across the network. For instance, these departments may serve as a hub for posting notices to patients as well as staff members of the hospital, hosting awareness creation activities, placing shared resources and propagating technology transfer projects.

**Table 2.** Betweeness and closeness centrality of departments.

| Departments | Betweeness Centrality | Departments | Closeness Centrality |
|---|---|---|---|
| Regular Laboratory | 6449 | Regular Laboratory | 0.71 |
| Emergency Laboratory1 | 5089 | Functional Diagnostic | 0.68 |
| Cardiology One | 3879 | Emergency Laboratory1 | 0.68 |
| Cardiology Two | 3819 | ICU | 0.63 |

Five communities of departments (see Fig. 2) with different proportion (C1 (Violet) = 37.4%, C2 (Green) = 27.6%, C3 (Azure) = 19%, C4 (Orange)= 14% and C5 (Forest) = 4%) were identified using modularity and community extraction algorithm proposed by Blondel et al [24].

The layout of the network of departments was produced in two steps: first, ForceAtlas2 algorithm [25] was applied to arrange nodes and edges. Second, Expansion algorithm was employed to scale up the network and make the layout more visible.

The average clustering coefficient equals to 0.62 which shows that there is strong interaction among departments within a module or community. These strongly connected departments can be reorganized and placed next to one another to maintain the geographical proximity between them. This may minimize time, cost and energy spend by the patients as well as health professionals to move from one department to another department. This may also improve the data and/or information exchange which, in turn, advances the coordination of departments as well as the effectiveness and efficiency of the diagnosis and treatment process in the center.

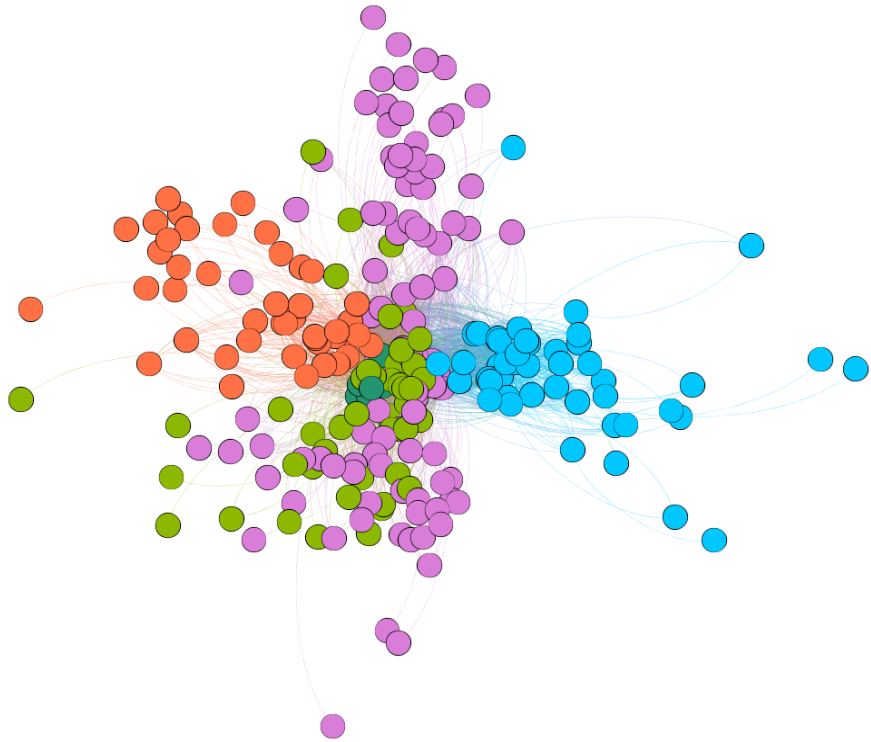

**Fig. 2.** Communities of departments.

## 3.2 Temporal Network of Departments

Due to different reasons such as advancement of technology or business process reengineering, the structure of network of departments and the role of departments may change over time. In this section, analysis of the structure of network of departments and the role of departments over seven years is discussed. The node size in the network increases from year to year (see Table 3) which may indicate an introduction of new departments (e.g., emergency laboratory 2, see Fig. 4) or technologies or a change in working process that allow existing departments to engage actively in the diagnosis and treatment process.

As the node size increases, the edge size also increases with positive correlation coefficient of 0.94 (see Figure 3). There is also positive correlation between edge size and average degree with a correlation coefficient of 0.76. Besides to this, modularity has a positive correlation with node size, edge size, Average Path Length (APL), and the number of strongly connected components.

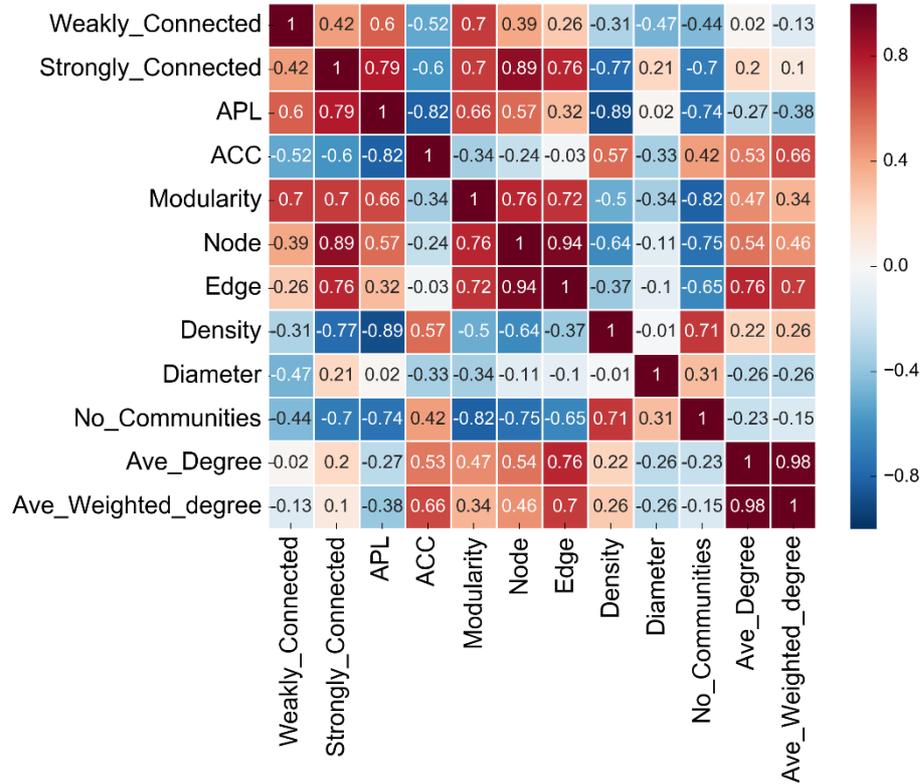

**Fig. 3.** Correlation coefficient of measurements of temporal network over years.

**Table 3.** Betweeness and closeness centrality of departments.

| Year | Weakly_ Con- nected | Strongly _ Con- nected | APL | ACC | Modu- larity | Node | Edge | Density | Diame- ter | Ave_ Degree | Ave_ Weighted _ degree |
|------|------|------|------|------|------|------|------|------|------|------|------|
| 2010 | 1 | 9 | 2.2 | 0.5 | 0.24 | 70 | 635 | 0.13 | 5 | 9 | 729 |
| 2011 | 1 | 5 | 2 | 0.62 | 0.16 | 77 | 1025 | 0.2 | 5 | 13 | 1440 |
| 2012 | 1 | 9 | 2.1 | 0.64 | 0.23 | 110 | 1282 | 0.1 | 4 | 12 | 1387 |
| 2013 | 1 | 3 | 2 | 0.7 | 0.32 | 101 | 1799 | 0.2 | 4 | 18 | 2275 |
| 2014 | 1 | 20 | 2.2 | 0.6 | 0.35 | 131 | 1909 | 0.1 | 5 | 15 | 1754 |
| 2015 | 1 | 28 | 2.2 | 0.55 | 0.34 | 170 | 2773 | 0.1 | 5 | 16 | 1897 |
| 2016 | 2 | 23 | 2.3 | 0.5 | 0.45 | 148 | 2060 | 0.1 | 4 | 14 | 1415 |

The strongly connected components (which refer to a sub-graphs in which all the departments are connected to each other by at least one path and have no connections with the rest of the graph [16]) also increases over time which was 9 in 2010 and became 23 in 2016. This may indicate that the probability of forming strong connection within community or clique of departments is higher than establishing connection between departments from different communities even though both node size and edge size grow.

In other words, as a new department added to the network and got connected with one of the departments in one of the communities in the network, it tends to strengthen the local interaction instead of forming new connection with other nodes from another region of the network. This is supported with a constant average path length that does not change over time and a strong positive correlation of modularity with node size as well as edge size (see Fig. 3 and Table 3).

There is negative correlation between APL and Average Clustering Coefficient (ACC). Density of the temporal network also has negative correlation with node size, APL and the number of strongly connected components. The number of communities in the network also has negative correlation with node size and edge size which may illustrate that the node size and edge size do not contribute to the number of communities.

When we see the role of departments over time, the betweenness centrality (see Fig. 4) of Emergency Laboratory1, Regular Laboratory and Functional Diagnostic departments demonstrate growing trend, whereas all cardiology departments display random walk.

On the other hand, the betweeness centrality of ICU1, ICU2 and Outpatient clinic stay constant over time. This may show that departments such as Emergency Laboratory1, Regular Laboratory and Functional Diagnostic departments which facilitate the diagnosis procedure are the back bone of the healthcare system that deliver medical for ACS patients.

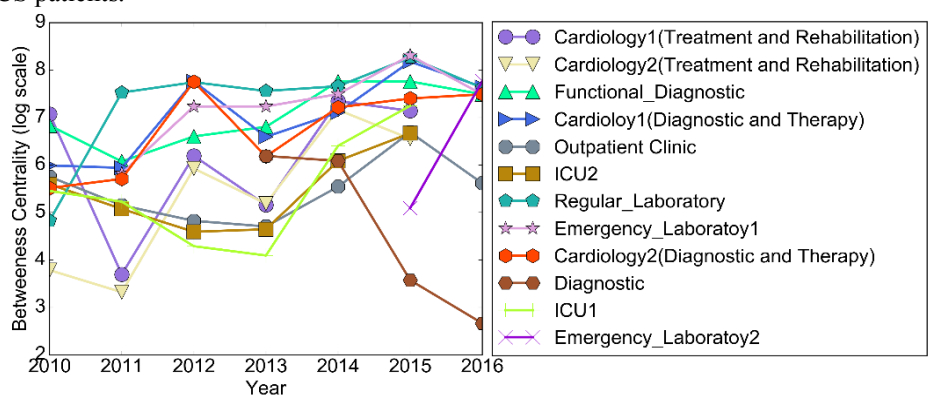

**Fig. 4.** Departments with high betweeness centrality over the course of seven years.

## 4 Conclusion and Future Work

In this article, data-driven static and temporal networks of departments are proposed to study the underlying structure of network of departments that deliver healthcare for patients, to identify the main departments and their role in the diagnosis and treatment process, to investigate evolution of role of departments over time, and to discover communities of departments.

Seven years', from 2010 to 2016, empirical data of 24902 Acute Coronary Syndrome (ACS) patients was employed to construct both static and temporal networks based on an episode-based transfer of patients.

As a result, we found out that Laboratory department, Emergency Laboratory1 department and Functional Diagnostic department are receiving more requests as well as sending more results to other departments. Five communities were discovered with an average clustering coefficient of 0.62. These departments are also fundamental in connecting communities of departments as they have high betweeness and closeness centrality in the network.

The results of this study may help hospital administration to effectively organize and manage the interaction among departments and sub-group of departments and maintain functionality, capacity and geographical proximity of departments according to their strategic positioning and role in the diagnosis and treatment process of ACS patients so that time, cost and energy could be saved and value-based healthcare could be achieved.

Finally, in the future, analysis of optimal arrangement of physical location of departments in a hospital will be conducted given degree, weighted degree, betweeness centrality, and closeness centrality measures. In addition, cost and time that take to transfer patients or information among departments will be considered. So that cost and time could be saved and patients' satisfaction could be improved.

Besides to this, how to achieve system approach in constructing holistic patient flow simulation, while maintaining the balance between the complexity and the simplicity of the model can be investigated using the results of this study.

## 5 Acknowledgement

This research is financially supported by The Russian Scientific Foundation, Agreement #17-15-01177.

## References

[1] Rosamond Hutt, "Economics of healthcare: which countries are getting it right? | World Economic Forum," 2016. [Online]. Available: https://www.weforum.org/agenda/2016/04/which-countries-have-the-most-cost-effective-healthcare/. [Accessed: 30-Aug-2018].

[2] W. A. Haseltine, "Aging Populations Will Challenge Healthcare Systems All Over The World," Forbes, 2018. [Online]. Available: https://www.forbes.com/sites/williamhaseltine/2018/04/02/aging-populations-will-challenge-healthcare-systems-all-over-the-world/#751ab4bd2cc3. [Accessed: 14-Dec-2018].

[3]    M. Papi, L. Pontecorvi, and R. Setola, "A new model for the length of stay of hospital patients," Health Care Manag. Sci., vol. 19, no. 1, pp. 58–65, Mar. 2016.

[4]    S. Chand, H. Moskowitz, J. B. Norris, S. Shade, and D. R. Willis, "Improving patient flow at an outpatient clinic: Study of sources of variability and improvement factors," Health Care Manag. Sci., vol. 12, no. 3, pp. 325–340, 2009.

[5]    N. D. Soulakis, M. B. Carson, Y. J. Lee, D. H. Schneider, C. T. Skeehan, and D. M. Scholtens, "Visualizing collaborative electronic health record usage for hospitalized patients with heart failure," J. Am. Med. Informatics Assoc., vol. 22, no. 2, pp. 299–311, 2015.

[6]    E. H. DuGoff, S. Fernandes-Taylor, G. E. Weissman, J. H. Huntley, and C. E. Pollack, "A scoping review of patient-sharing network studies using administrative data," Transl. Behav. Med., vol. 8, no. 4, pp. 598–625, Jul. 2018.

[7]    A. Suraj, S. Kundu, and W. Norcross, "Diagnosis of Acute Coronary Syndrome," American Family Physician, 2005. [Online]. Available: https://www.aafp.org/afp/2005/0701/p119.pdf. [Accessed: 26-May-2018].

[8]    "Acute Coronary Syndrome," American Heart Association, 2017. [Online]. Available: http://www.heart.org/HEARTORG/Conditions/HeartAttack/AboutHeartAttacks/Acute-Coronary-Syndrome_UCM_428752_Article.jsp#.WwgvDi5uaUk. [Accessed: 25-May-2018].

[9]    WHO, "Cardiovascular diseases (CVDs)," 2017. [Online]. Available: http://www.who.int/news-room/fact-sheets/detail/cardiovascular-diseases-(cvds). [Accessed: 24-Aug-2018].

[10]    P. Bhattacharjee and P. Kumar Ray, "Patient flow modelling and performance analysis of healthcare delivery processes in hospitals: A review and reflections," Comput. Ind. Eng., vol. 78, pp. 299–312, 2014.

[11]    E. Bullmore and O. Sporns, "Complex brain networks: graph theoretical analysis of structural and functional systems," Nat. Rev. Neurosci., vol. 10, no. 3, pp. 186–198, Mar. 2009.

[12]    J.-P. Onnela, A. J. O'Malley, N. L. Keating, and B. E. Landon, "Comparison of physician networks constructed from thresholded ties versus shared clinical episodes," Appl. Netw. Sci., vol. 3, no. 1, p. 28, Dec. 2018.

[13]    Á. Rebuge and D. R. Ferreira, "Business process analysis in healthcare environments: A methodology based on process mining," Inf. Syst., vol. 37, no. 2, pp. 99–116, Apr. 2012.

[14]    E. Rojas, J. Munoz-Gama, M. Sepúlveda, and D. Capurro, "Process mining in healthcare: A literature review," J. Biomed. Inform., vol. 61, pp. 224–236, Jun. 2016.

[15]    "Chapter 4, Emerging Trends in Care Coordination Measurement | Agency for Healthcare Research and Quality." [Online]. Available: https://www.ahrq.gov/professionals/prevention-chronic-care/improve/coordination/atlas2014/chapter4.html#social. [Accessed: 28-Aug-2018].

[16]    S. Tabassum, F. S. F. Pereira, S. Fernandes, and J. Gama, "Social network analysis: An overview," Wiley Interdiscip. Rev. Data Min. Knowl. Discov., vol. 8, no. 5, pp. 1–21, Sep. 2018.

[17]    A. G. Dunn and J. I. Westbrook, "Interpreting social network metrics in healthcare organisations: A review and guide to validating small networks," Soc. Sci. Med., vol. 72, no. 7, pp. 1064–1068, Apr. 2011.

[18]    L. Benhiba, A. Loutfi, M. Abdou, and J. Idrissi, "A Classification of Healthcare Social Network Analysis Applications," in HEALTHINF 2017 - 10th International Conference on Health Informatics, 2017, pp. 147–158.

[19]    D. Chambers, P. Wilson, C. Thompson, and M. Harden, "Social network analysis in healthcare settings: a systematic scoping review.," PLoS One, vol. 7, no. 8, p. e41911, 2012.

[20]   S.-H. Bae, A. Nikolaev, J. Y. Seo, and J. Castner, "Health care provider social network analysis: A systematic review," Nurs. Outlook, vol. 63, no. 5, pp. 566–584, Sep. 2015.

[21]   F. Wang, U. Srinivasan, S. Uddin, and S. Chawla, "Application of Network Analysis on Healthcare," in 2014 IEEE/ACM International Conference on Advances in Social Networks Analysis and Mining (ASONAM 2014), 2014, no. Asonam, pp. 596–603.

[22]   A. De Brún and E. McAuliffe, "Social Network Analysis as a Methodological Approach to Explore Health Systems: A Case Study Exploring Support among Senior Managers/Executives in a Hospital Network.," Int. J. Environ. Res. Public Health, vol. 15, no. 3, Mar. 2018.

[23]   "Gephi - The Open Graph Viz Platform." [Online]. Available: https://gephi.org/. [Accessed: 23-Jan-2019].

[24]   A. Iop, V. D. Blondel, J.-L. Guillaume, R. Lambiotte, and E. Lefebvre, "Fast unfolding of communities in large networks," J. Stat. Mech, p. 10008, 2008.

[25]   M. Jacomy, T. Venturini, S. Heymann, and M. Bastian, "ForceAtlas2, a Continuous Graph Layout Algorithm for Handy Network Visualization Designed for the Gephi Software," PLoS One, vol. 9, no. 6, p. e98679, Jun. 2014.